\documentclass{article}

\usepackage{spconf,amsmath,graphicx}
\usepackage{multirow, diagbox}
\usepackage[hidelinks]{hyperref} 

\title{Evaluating Feature Attribution Methods for Electrocardiogram}

\name{Jangwon Suh$^{\star}$ \qquad Jimyeong Kim$^{\star}$ \qquad Euna Jung$^{\star}$ \qquad Wonjong Rhee$^{\star \dagger}$ \thanks{This work was supported by the Korea Medical Device Development Fund grant funded by the Korea government (the Ministry of Science and ICT, the Ministry of Trade, Industry and Energy, the Ministry of Health \& Welfare, the Ministry of Food and Drug Safety) (Project Number: HI20C1662, 1711138358, KMDF PR 20200901 0173).}}

\address{$^{\star}$ Department of Intelligence and Information, Seoul National University \\
$^{\dagger}$ Interdisciplinary Program in Artificial Intelligence, Seoul National University}

\begin{document}
%
\maketitle
\begin{abstract}

The performance of cardiac arrhythmia detection with electrocardiograms~(ECGs) has been considerably improved since the introduction of deep learning models. In practice, the high performance alone is not sufficient and a proper explanation is also required. Recently, researchers have started adopting feature attribution methods to address this requirement, but it has been unclear which of the methods are appropriate for ECG. In this work, we identify and customize three evaluation metrics for feature attribution methods based on the characteristics of ECG: localization score, pointing game, and degradation score. Using the three evaluation metrics, we evaluate and analyze eleven widely-used feature attribution methods. We find that some of the feature attribution methods are much more adequate for explaining ECG, where Grad- CAM outperforms the second-best method by a large margin.

\end{abstract}
%

\section{Introduction}
\label{sec:intro}
For electrocardiograms~(ECGs), deep learning models have achieved a near cardiologist-level performance on arrhythmia detection and classification~\cite{rajpurkar2017cardiologist}. The models, however, are often considered to be a “black box” because the decision-making process is challenging to explain. For providing an explanation in the image domain, feature attribution has become a popular technique where the relevant features in the input can be highlighted~\cite{gevaert2022evaluating}. It is also believed to be useful for ECG domain because a cardiac abnormality is usually declared based on a specific area of an ECG. Therefore, its usage has been consistently increasing. However, there has been no clear analysis on which feature attribution methods are adequate for ECG~\cite{neves2021interpretable}.

In this study, we examine eleven feature attribution methods that are widely used in the image domain~\cite{gevaert2022evaluating, schlegel2019towards}, and determine which methods are suitable for ECG. We propose three evaluation metrics identified and customized by considering ECG-specific characteristics. Using the three metrics, we evaluate the eleven attribution methods.
Our main contributions are:
(1) We propose three metrics for evaluating feature attribution methods on ECG - \textbf{localization score}, \textbf{pointing game}, and \textbf{degradation score}.
(2) We evaluate the eleven feature attribution methods with the proposed metrics and provide a quantitative analysis.
(3) We find that Grad-CAM is the most appropriate for ECG data. Our experiments can be easily reproduced and transferred to other experimental settings, because they are implemented with Captum~\cite{kokhlikyan2020captum} that is an open-source library for model interpretability and because we use a publicly available ECG dataset. Our code is available at {\small \url{https://github.com/SNU-DRL/Attribution-ECG}}.


\section{Backgrounds}
\label{sec:backgrounds}

\subsection{Feature Attribution Methods}
\label{ssec:feature_attribution_methods}

Feature attribution is a technique that measures the importance of each pixel in an image for a given prediction of a model. Because feature attribution values can be easily visualized as a heatmap showing which pixels are important, it has become popular in the image domain~\cite{gevaert2022evaluating}. Feature attribution methods can be categorized according to the mechanisms they use~\cite{gevaert2022evaluating, rao2022towards}.
\textit{Backpropagation-based} methods use gradients of a model computed by the backpropagation process to measure the importance of each element in the input. This group includes Saliency~\cite{simonyan2013deep}, Input$\times$Gradient~\cite{shrikumar2017learning}, Guided Backprop~\cite{springenberg2014striving}, Integrated Gradients~\cite{sundararajan2017axiomatic}, DeepLIFT~\cite{shrikumar2017learning}, DeepSHAP~\cite{lundberg2017unified} and LRP \cite{bach2015pixel}.
\textit{Perturbation-based} methods measure feature attributions by tracking the model outputs for the given perturbed inputs. This group includes LIME~\cite{ribeiro2016should} and KernelSHAP~\cite{lundberg2017unified}.
\textit{Activation-based} methods use activation maps where the importance for the target class is computed. This group includes Grad-CAM~\cite{selvaraju2017grad}. We also classify Guided Grad-CAM~\cite{selvaraju2017grad}, which combines Grad-CAM and Guided Backprop, as an activation-based.

\vspace*{-3pt}

\subsection{Evaluation Metrics for Feature Attribution Methods}
\label{ssec:evaluation_metrics_for_feature_attribution_methods}

A series of studies have proposed evaluation metrics to analyze and compare feature attribution in the context of computer vision research~\cite{gevaert2022evaluating}. Evaluation metrics examine whether each attribution method explains a model's prediction with a high fidelity. They can be categorized mainly into two groups: \textit{localization-based metrics} and \textit{perturbation-based metrics}.

Localization-based metrics~\cite{simonyan2013deep, cao2015look, schulz2020restricting, zhang2018top,fong2017interpretable,bohle2021convolutional} assess how successfully a feature attribution method localizes the input's class-discriminative features. To evaluate feature attribution methods using localization-based metrics, we need a dataset with additional ground-truth information on which area of an input is closely related to the label. For this reason, researchers in the image domain have conducted experiments with special datasets containing object bounding boxes or have constructed image grids~\cite{rao2022towards}. As localization-based metrics utilize human-made labels, they reflect human knowledge where a model should attend to.

Perturbation-based metrics~\cite{schulz2020restricting, samek2016evaluating, ancona2017towards} assess feature attribution methods by perturbing an input and measuring how much the prediction of a model is affected. If an attribution method is effective, its heatmap should match the areas that are sensitive to the perturbation. An advantage of perturbation-based metrics is that they do not require additional ground-truth information. Therefore, perturbation-based metrics are applicable to a wider variety of datasets than localization-based metrics.

\subsection{Evaluating Feature Attribution Methods for ECG}
\label{ssec:evaluating_feature_attribution_methods_for_ecg}

Feature attribution methods and evaluation metrics for them have been mainly developed in the computer vision research.
The results from \cite{gevaert2022evaluating} imply that the performance of attribution methods should be measured for each specific use case. Thus, to apply feature attribution methods to ECG domain, such methods should be verified using ECG data and tasks.
There are a limited number of studies that focus on evaluating feature attribution methods using ECG data~\cite{neves2021interpretable, schlegel2019towards, turbe2022interprettime}. However, the evaluation metrics used in the previous studies are designed for general time-series data based on shapelets~\cite{neves2021interpretable} or perturbation-based metrics~\cite{neves2021interpretable, schlegel2019towards, turbe2022interprettime}. Moreover, the number of feature attribution methods studied in the three studies is at most six. 

A sound evaluation can be achieved by concurrently evaluating additional feature attribution methods with customized metrics that reflect ECG-specific characteristics.
In this study, we propose two localization-based metrics that utilize the pseudo-periodic structure of an ECG signal and one perturbation-based metric that jointly considers the performance drop under the increasing and decreasing order of attribution values.
We compare eleven feature attribution methods with our proposed evaluation metrics, which is roughly twice of 
the previous studies.

\begin{figure}[htb]
  \centering
  \includegraphics[width=8.5cm]{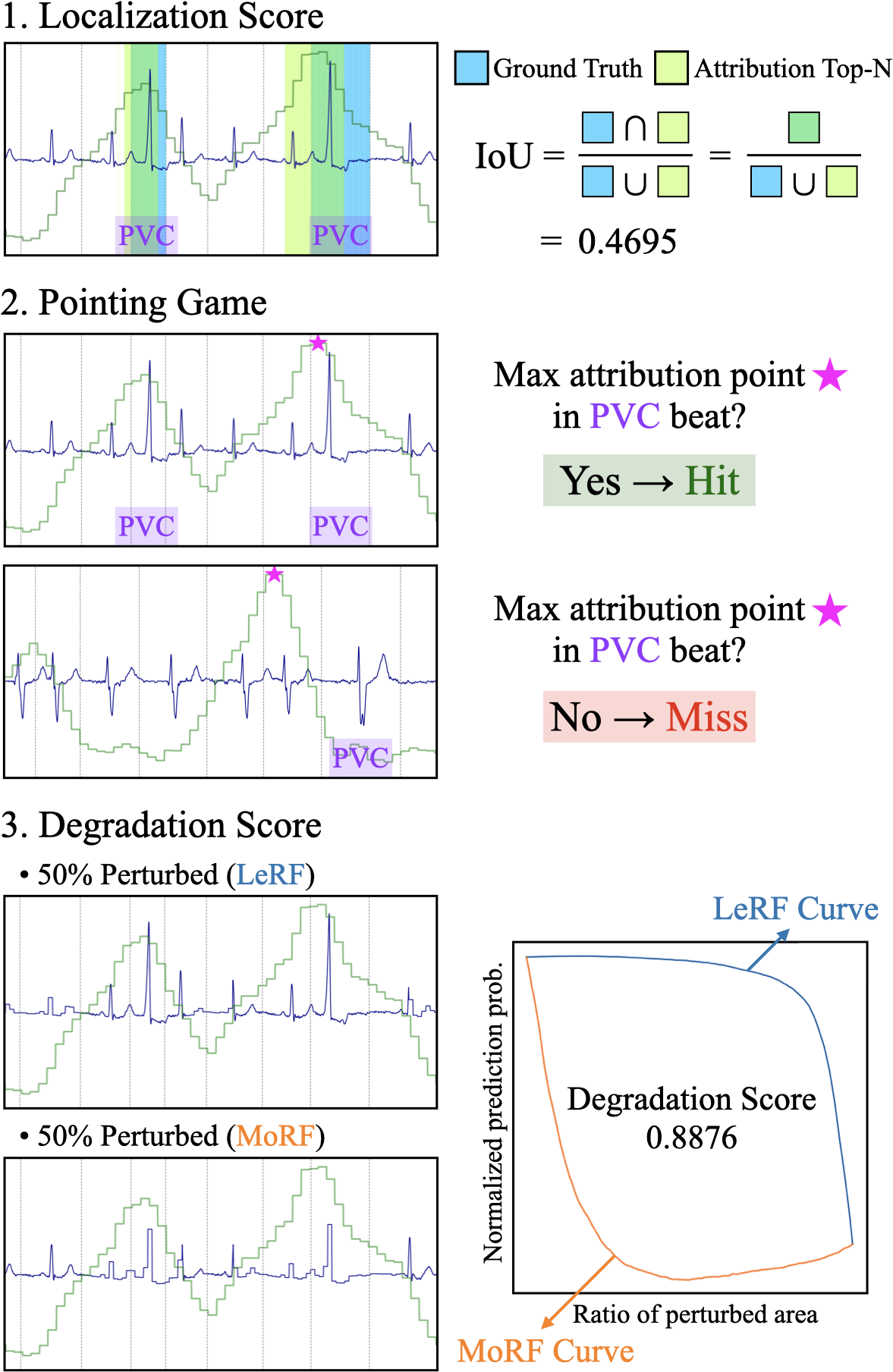}
\caption{Overview of our proposed three evaluation metrics. ECG signals~(blue line) of PVC are visualized with Grad-CAM attributions~(green line). As the localization score and the pointing game need information on where abnormal beats are, we additionally mark PVC beats in each example.}
\label{fig:evaluation-metrics}
\end{figure}

\section{Proposed metrics}
\label{sec:proposed-metrics}
\begin{figure}[htb]
  \centering
  \includegraphics[width=8.5cm]{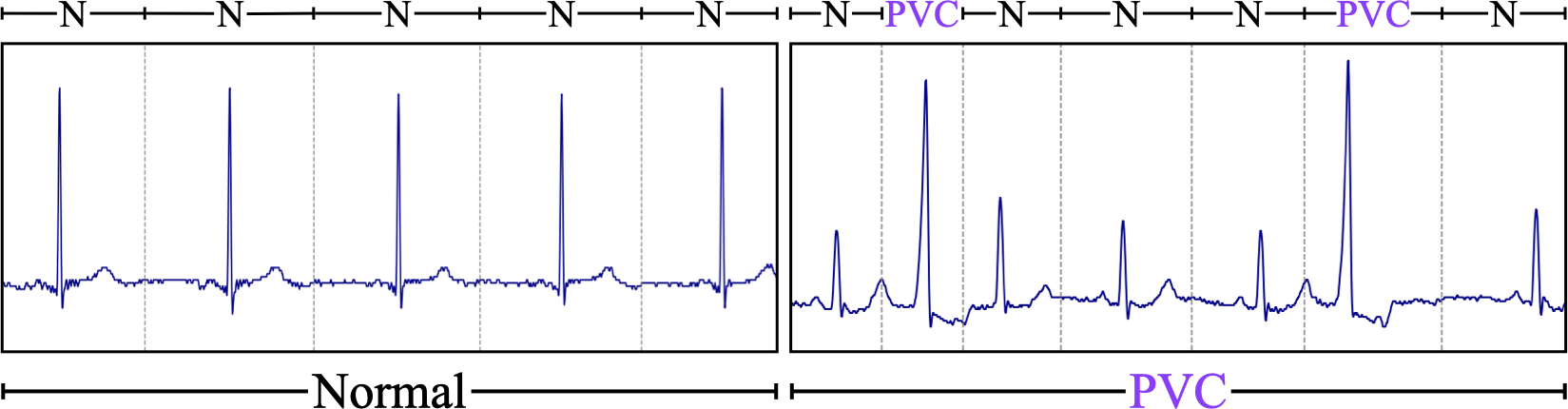}
\caption{Illustration of constructing the label of an ECG example that consists of multiple beats. An example consisting of all normal beats is labeled as normal; An example including any PVC beats is labeled as PVC.}
\vspace{-5pt}
\label{fig:beat_level_span_level}
\end{figure}

\vspace{-5pt}
We propose three metrics to evaluate feature attribution methods for ECG. The overview of our proposed evaluation metrics is shown in Figure \ref{fig:evaluation-metrics}.
Localization-based metrics, \textit{Localization Score} and \textit{Pointing Game}, need extra information where the true cardiac abnormalities lie in an example.
To implement those metrics, we propose using an ECG dataset with labels for every beat.
Since we mostly focus on detecting abnormal beats out of normal beats, we could regard the label of an example as the label of any abnormal beats present if the example has any such beats, as shown in Figure \ref{fig:beat_level_span_level}.
We have the information on where the abnormal beats are, so we can apply localization-based metrics to ECG domain.
This is similar to image datasets containing object bounding boxes being used when evaluating with localization-based metrics in the image domain.
On the other hand, Perturbation-based metric, \textit{Degradation Score}, can be applied to ECG dataset with any type of labels - labels for a beat or for a specific span of ECG.

\subsection{Localization Score}
\label{ssec:localization-score}
\textbf{Localization Score} is a metric that measures the overlap between the area with high attribution values and the ground truth area~\cite{simonyan2013deep, cao2015look, schulz2020restricting}. Considering the characteristics of the ECG data, we propose the following metric for calculating the localization score. If the abnormal beats of an example contain $n$ data points, we select top-$n$ data points with the highest attribution values. Then we calculate the Intersection-over-Union~(IoU) between the selected $n$ data points and the $n$ data points in abnormal beats. IoU between two sets $A$ and $B$ is defined as follows:
\begin{equation}\label{iou}
\vspace{-7pt}
\text{IoU}(A,B) = \frac{|A \cap B|}{|A \cup B|}.
\vspace{-1pt}
\end{equation}

\subsection{Pointing Game}
\label{ssec:pointing_game}
\textbf{Pointing Game} is a metric that measures if a data point with the maximum attribution value correctly points to the ground truth area~\cite{zhang2018top,fong2017interpretable,bohle2021convolutional}. This metric is different from the localization score in that the pointing game does not require highlighting the full extent of an object and only focuses on the maximum attribution point~\cite{zhang2018top}. We measure the pointing game accuracy as follows: If a data point with the highest attribution value falls on one of the abnormal beats, it counts as a hit; otherwise, it counts as a miss. Then we calculate the accuracy as described in \cite{zhang2018top}.
\begin{equation}\label{pointing_game_accuracy}
\vspace{-7pt}
    \text{Accuracy} = \frac{\#Hits}{\#Hits + \#Misses}
\vspace{-1pt}
\end{equation}

\subsection{Degradation Score}
\label{ssec:degradation_score}

\textbf{Degradation Score} is a metric that examines how a model's prediction degrades under cumulative perturbation of an original input, suggested by \cite{schulz2020restricting}.
Perturbations can be applied to small-sized windows of the input in the order of decreasing attribution values (Most Relevant removed First, MoRF) \cite{samek2016evaluating} or of increasing attribution values (Least Relevant removed First, LeRF)~\cite{ancona2017unified}. The degradation score combines both perturbation processes.
We scale the model's output probability for a correct class of an original example to one and the probability of a fully perturbed example to zero.
Then we plot MoRF and LeRF curves as shown in the lower right side of Figure \ref{fig:evaluation-metrics} and compute the integral between the two curves. As a perturbation, we use a mean operation; we fill a selected window with the mean value of the window. We select the mean operation since it can erase detailed wave information while preserving the typical shape of an ECG signal.

\section{Experiments}
\label{sec:experiments}

\begin{table}[b]
\centering
    \begin{tabular}{c|r}
    \begin{tabular}[c]{@{}c@{}}Model Architecture \end{tabular}                 & \multicolumn{1}{c}{\begin{tabular}[c]{@{}c@{}} Accuracy (\%)\end{tabular}} \\ \hline
    ConvResNet (48 layers)~\cite{rajpurkar2017cardiologist, tan2021icentia11k}     & 0.85 ± 0.03                                                                                     \\
    Basic ConvNet (5 layers)~\cite{tan2021icentia11k}                              & 0.80 ± 0.01                                                                                     \\
    \begin{tabular}[c]{@{}c@{}}Modified ResNet (18 layers) (Ours)\end{tabular} & \textbf{0.90 ± 0.05}
        \\
    \end{tabular}
    \caption{Classification accuracy on Icentia11K beat classification task. The results of ConvResNet and Basic ConvNet are from \cite{tan2021icentia11k} for comparison.}
    \label{tab:beat_classification_accuracy}
\end{table}

\begin{table*}[t]
  \centering
\begin{tabular}{cc|>{\centering}p{0.1\textwidth}>{\centering}p{0.1\textwidth}>{\centering}p{0.1\textwidth}|c}
\multicolumn{2}{c|}{\diagbox{Attribution Methods}{Evaluation Metrics}} & \multicolumn{1}{c}{\begin{tabular}[c]{@{}c@{}}Localization\\ Score\end{tabular}} & \multicolumn{1}{c}{\begin{tabular}[c]{@{}c@{}}Pointing\\ Game\end{tabular}} & \multicolumn{1}{c|}{\begin{tabular}[c]{@{}c@{}}Degradation\\ Score\end{tabular}} & \multicolumn{1}{c}{Average}  \\ 
\hline
\multicolumn{2}{c|}{Random (baseline)} & 0.1087 & 0.1857 & 0.0037 & 0.0094                       \\ 
\hline
\multirow{7}{*}{Backpropagation-based}     & Saliency$^\circ$ \cite{simonyan2013deep} & 0.1899 & 0.3646 & 0.3460  & 0.3002 \\
                                    & Input$\times$Gradient$^\circ$ \cite{shrikumar2017learning} & 0.2178 & 0.5735 & 0.8751 & 0.5555 \\
                                    & Guided Backprop$^\circ$ \cite{springenberg2014striving} & 0.1899 & 0.3646 & 0.3460 & 0.3002 \\
                                    & Integrated Gradients$^\circ$ \cite{sundararajan2017axiomatic} & 0.2236 & 0.6837 & 0.8852 & 0.5825 \\
                                    & DeepLIFT$^\circ$ \cite{shrikumar2017learning} & 0.2193 & 0.5901 & 0.8791 & 0.5628 \\
                                    & DeepSHAP$^\circ$ \cite{lundberg2017unified} & 0.2134 & 0.5557 & 0.8827 & 0.5506 \\
                                    & LRP$^\circ$ \cite{bach2015pixel} & 0.1375 & 0.2250 & 0.3844 & 0.2490 \\ 
\hline
\multirow{2}{*}{Perturbation-based} & LIME$^\bullet$ \cite{ribeiro2016should} & 0.1214 & 0.3398 & 0.4230 & 0.2497 \\
                                    & KernelSHAP$^\bullet$ \cite{lundberg2017unified} & 0.1155 & 0.3977 & 0.2367 & 0.2500 \\ 
\hline
\multirow{2}{*}{Activation-based}   & Grad-CAM$^\bullet$ \cite{selvaraju2017grad} & \textbf{0.4848} & \textbf{0.7048} & \textbf{0.8876} & \textbf{0.6924} \\
                                    & Guided Grad-CAM$^\circ$ \cite{selvaraju2017grad} & 0.3148 & 0.5710 & 0.6943 & 0.5267                      
\end{tabular}
    \vspace{-2pt}
    \caption{Evaluation of feature attribution methods with the proposed evaluation metrics. 
    A method with a black dot $^{\bullet}$ or a white dot $^{\circ}$ denotes the use of raw values or absolute values, respectively, for the attribution calculation. Between the two options, only the better performing result is reported}
    \vspace{-10pt}
    \label{tab:results}
\end{table*}

\subsection{Experimental Setup}

\textbf{Dataset.}
\label{ssec:dataset}
We experimented with the Icentia11K dataset~\cite{tan2021icentia11k, tan2022physionet}. Icentia11K is one of the largest public ECG datasets containing 11 thousand patients and 2 billion labeled beats from a single-lead ECG device. We mostly followed the experimental settings of the original Icentia11K beat classification task~\cite{tan2021icentia11k}. The goal of the beat classification task is to predict if an example consists of all normal beats or contains at least one premature atrial contraction~(PAC) or premature ventricular contraction~(PVC). Each ECG signal has 2049 data points that correspond to about 8 seconds, as the sampling rate of the dataset is 250Hz.
We use R-peak locations provided in the dataset to extract beats by using the midpoint of two adjacent R-peaks as the endpoint of a beat as done in \cite{jones2020improving}.

\textbf{Training and Evaluation.}
\label{ssec:training_and_evaluation}
Each of the training and test sets contain 6000 examples - 2000 for normal, PAC, and PVC classes respectively.
When evaluating feature attribution methods with the proposed metrics, we only used test examples correctly predicted as PAC or PVC with over 90\% probability.
The reason for using only PAC and PVC examples for evaluation is that the localization-based metrics cannot be measured on an example labeled as normal since it consists of all normal beats, and all beats are true localization targets.
Moreover, the reason for using only correctly predicted examples with high probability is to ensure that the low evaluation result of feature attribution methods is attributable to the attribution method and not the classification model's poor performance~\cite{bohle2021convolutional}.
The total number of examples used to compute evaluation metrics is around 3272 on average.

\textbf{Implementation Details.}
\label{ssec:implementation_details}
Since the original ResNet~\cite{he2016deep} was created for 2D images, we build a modified 18-layer ResNet with $1$x$7$-sized kernels for designed for 1D ECG signals.
Each ECG signal of 2049 points is standardized to have zero mean and one standard deviation. The model is trained with cross-entropy loss and Adam optimizer. We set learning rate to 5e-4, batch size to 128, weight decay to 1e-7, and training epochs to 50. The classification accuracy of our model on the test set is reported in Table \ref{tab:beat_classification_accuracy}, which exceeds that of previous models evaluated on the same dataset.
For evaluating the degradation score, we set the perturbation window size to 16.
All results are the average of 5 repeated experiments with different dataset splits and model initializations.

\subsection{Evaluation Result and Discussion} 
\label{ssec:evaluation and comparison}

\begin{figure}[htb]
  \centering
  \includegraphics[width=8.5cm]{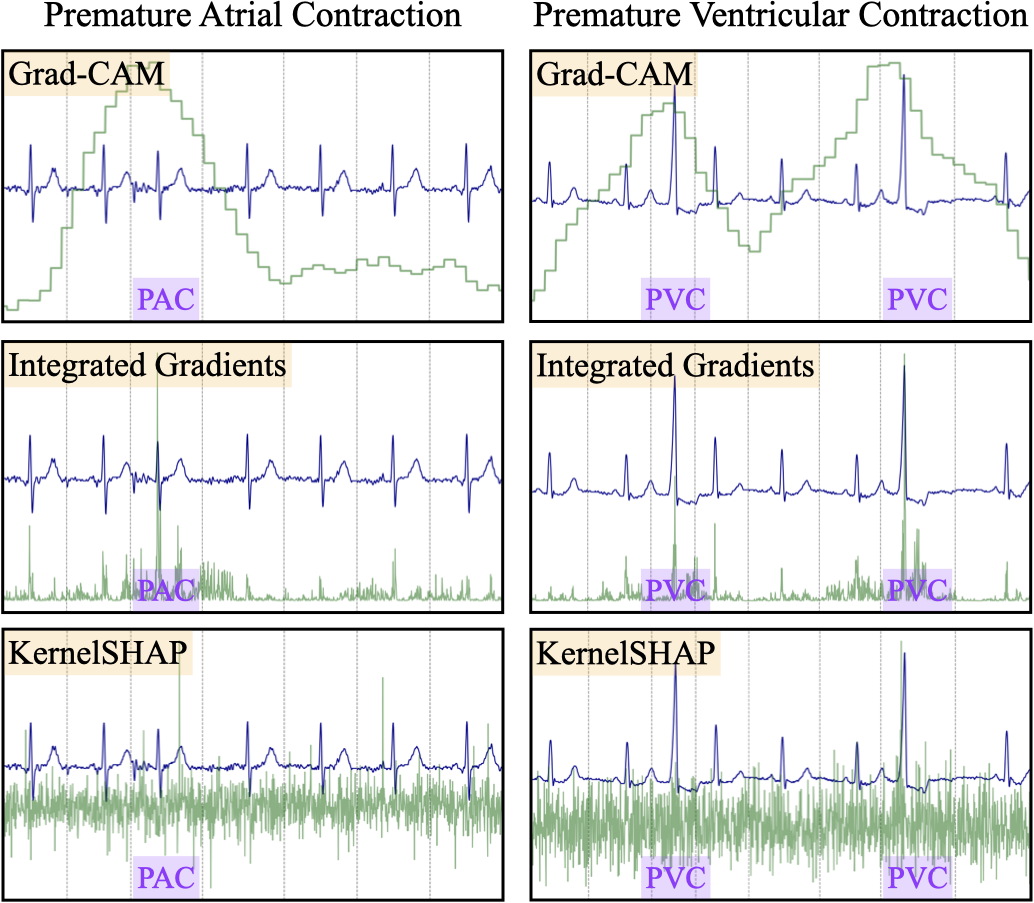}
\caption{
Visualization of ECG signals~(blue line) with Grad-CAM, Integrated Gradients, and KernelSHAP attributions~(green line).
Grad-CAM clearly shows a strong relationship with the ground-truth beats of PAC and PVC.
}
\label{fig:visualziation_feature_attribution_methods}
\end{figure}

The results are shown in Table~\ref{tab:results}. 
Among the eleven attribution methods, Grad-CAM achieves the best performance across all three evaluation metrics. For the localization score, the two activation-based methods, including Grad-CAM, significantly outperform all of the backpropagation-based and perturbation-based methods. For the pointing game, only Integrated Gradients performs comparably with Grad-CAM. For the degradation score, four of the backpropagation-based methods perform comparably with Grad-CAM. Overall, the average score of Grad-CAM is outstanding where it is 18\% better than the second-best method, Integrated Gradients. The visualization is provided in Figure~\ref{fig:visualziation_feature_attribution_methods}, where the three best methods from each category are shown.

An extensive evaluation of feature attribution methods in the image domain can be found in \cite{gevaert2022evaluating}. Compared to the image domain, ECG domain turns out to be distinct in a few ways. For instance, the two perturbation-based methods, LIME and KernelSHAP, perform reasonably well in the image domain but poorly in the ECG domain. 
DeepSHAP, which is known to achieve a competitive performance to Grad-CAM on ImageNet~\cite{gevaert2022evaluating}, also performs poorly in the ECG domain except for the degradation score.
Grad-CAM, however, turns out to perform well for both domains. As confirmed in \cite{gevaert2022evaluating}, different attribution metrics seem to measure different underlying concepts. For ECG domain, we can conclude that Grad-CAM is an appropriate feature attribution method that measures underlying concepts that are strongly relevant to cardiac arrhythmia detection.

\section{Conclusion}
\label{sec:conclusion}

We have identified and customized three evaluation metrics for ECG and assessed eleven feature attribution methods using the metrics. The experiment results show that Grad-CAM outperforms the other methods including the ones that are known to perform well in the image domain. A possible future work is to design a new attribution method that specifically utilizes the characteristics of ECG.
A limitation of our work is that we have considered only PAC and PVC with Icentia11K dataset.

\vfill\pagebreak



\let\oldbibliography\thebibliography
\renewcommand{\thebibliography}[1]{\oldbibliography{#1}
\setlength{\itemsep}{-2pt}} 

\small{\bibliographystyle{IEEEbib}}
\bibliography{strings,refs}

\begin{thebibliography}{10}

\bibitem{rajpurkar2017cardiologist}
Pranav Rajpurkar, Awni~Y Hannun, Masoumeh Haghpanahi, Codie Bourn, and Andrew~Y
  Ng,
\newblock ``Cardiologist-level arrhythmia detection with convolutional neural
  networks,''
\newblock {\em arXiv preprint arXiv:1707.01836}, 2017.

\bibitem{gevaert2022evaluating}
Arne Gevaert, Axel-Jan Rousseau, Thijs Becker, Dirk Valkenborg, Tijl De~Bie,
  and Yvan Saeys,
\newblock ``Evaluating feature attribution methods in the image domain,''
\newblock {\em arXiv preprint arXiv:2202.12270}, 2022.

\bibitem{neves2021interpretable}
In{\^e}s Neves, Duarte Folgado, Sara Santos, Mar{\'\i}lia Barandas, Andrea
  Campagner, Luca Ronzio, Federico Cabitza, and Hugo Gamboa,
\newblock ``Interpretable heartbeat classification using local model-agnostic
  explanations on ecgs,''
\newblock {\em Computers in Biology and Medicine}, vol. 133, pp. 104393, 2021.

\bibitem{schlegel2019towards}
Udo Schlegel, Hiba Arnout, Mennatallah El-Assady, Daniela Oelke, and Daniel~A
  Keim,
\newblock ``Towards a rigorous evaluation of xai methods on time series,''
\newblock in {\em 2019 IEEE/CVF International Conference on Computer Vision
  Workshop (ICCVW)}. IEEE, 2019, pp. 4197--4201.

\bibitem{kokhlikyan2020captum}
Narine Kokhlikyan, Vivek Miglani, Miguel Martin, Edward Wang, Bilal Alsallakh,
  Jonathan Reynolds, Alexander Melnikov, Natalia Kliushkina, Carlos Araya, Siqi
  Yan, and Orion Reblitz-Richardson,
\newblock ``Captum: A unified and generic model interpretability library for
  pytorch,'' 2020.

\bibitem{rao2022towards}
Sukrut Rao, Moritz B{\"o}hle, and Bernt Schiele,
\newblock ``Towards better understanding attribution methods,''
\newblock in {\em Proceedings of the IEEE/CVF Conference on Computer Vision and
  Pattern Recognition}, 2022, pp. 10223--10232.

\bibitem{simonyan2013deep}
Karen Simonyan, Andrea Vedaldi, and Andrew Zisserman,
\newblock ``Deep inside convolutional networks: Visualising image
  classification models and saliency maps,''
\newblock {\em arXiv preprint arXiv:1312.6034}, 2013.

\bibitem{shrikumar2017learning}
Avanti Shrikumar, Peyton Greenside, and Anshul Kundaje,
\newblock ``Learning important features through propagating activation
  differences,''
\newblock in {\em International conference on machine learning}. PMLR, 2017,
  pp. 3145--3153.

\bibitem{springenberg2014striving}
Jost~Tobias Springenberg, Alexey Dosovitskiy, Thomas Brox, and Martin
  Riedmiller,
\newblock ``Striving for simplicity: The all convolutional net,''
\newblock {\em arXiv preprint arXiv:1412.6806}, 2014.

\bibitem{sundararajan2017axiomatic}
Mukund Sundararajan, Ankur Taly, and Qiqi Yan,
\newblock ``Axiomatic attribution for deep networks,''
\newblock in {\em International conference on machine learning}. PMLR, 2017,
  pp. 3319--3328.

\bibitem{lundberg2017unified}
Scott~M Lundberg and Su-In Lee,
\newblock ``A unified approach to interpreting model predictions,''
\newblock {\em Advances in neural information processing systems}, vol. 30,
  2017.

\bibitem{bach2015pixel}
Sebastian Bach, Alexander Binder, Gr{\'e}goire Montavon, Frederick Klauschen,
  Klaus-Robert M{\"u}ller, and Wojciech Samek,
\newblock ``On pixel-wise explanations for non-linear classifier decisions by
  layer-wise relevance propagation,''
\newblock {\em PloS one}, vol. 10, no. 7, pp. e0130140, 2015.

\bibitem{ribeiro2016should}
Marco~Tulio Ribeiro, Sameer Singh, and Carlos Guestrin,
\newblock ``" why should i trust you?" explaining the predictions of any
  classifier,''
\newblock in {\em Proceedings of the 22nd ACM SIGKDD international conference
  on knowledge discovery and data mining}, 2016, pp. 1135--1144.

\bibitem{selvaraju2017grad}
Ramprasaath~R Selvaraju, Michael Cogswell, Abhishek Das, Ramakrishna Vedantam,
  Devi Parikh, and Dhruv Batra,
\newblock ``Grad-cam: Visual explanations from deep networks via gradient-based
  localization,''
\newblock in {\em Proceedings of the IEEE international conference on computer
  vision}, 2017, pp. 618--626.

\bibitem{cao2015look}
Chunshui Cao, Xianming Liu, Yi~Yang, Yinan Yu, Jiang Wang, Zilei Wang, Yongzhen
  Huang, Liang Wang, Chang Huang, Wei Xu, et~al.,
\newblock ``Look and think twice: Capturing top-down visual attention with
  feedback convolutional neural networks,''
\newblock in {\em Proceedings of the IEEE international conference on computer
  vision}, 2015, pp. 2956--2964.

\bibitem{schulz2020restricting}
Karl Schulz, Leon Sixt, Federico Tombari, and Tim Landgraf,
\newblock ``Restricting the flow: Information bottlenecks for attribution,''
\newblock in {\em International Conference on Learning Representations}, 2020.

\bibitem{zhang2018top}
Jianming Zhang, Sarah~Adel Bargal, Zhe Lin, Jonathan Brandt, Xiaohui Shen, and
  Stan Sclaroff,
\newblock ``Top-down neural attention by excitation backprop,''
\newblock {\em International Journal of Computer Vision}, vol. 126, no. 10, pp.
  1084--1102, 2018.

\bibitem{fong2017interpretable}
Ruth~C Fong and Andrea Vedaldi,
\newblock ``Interpretable explanations of black boxes by meaningful
  perturbation,''
\newblock in {\em Proceedings of the IEEE international conference on computer
  vision}, 2017, pp. 3429--3437.

\bibitem{bohle2021convolutional}
Moritz Bohle, Mario Fritz, and Bernt Schiele,
\newblock ``Convolutional dynamic alignment networks for interpretable
  classifications,''
\newblock in {\em Proceedings of the IEEE/CVF Conference on Computer Vision and
  Pattern Recognition}, 2021, pp. 10029--10038.

\bibitem{samek2016evaluating}
Wojciech Samek, Alexander Binder, Gr{\'e}goire Montavon, Sebastian Lapuschkin,
  and Klaus-Robert M{\"u}ller,
\newblock ``Evaluating the visualization of what a deep neural network has
  learned,''
\newblock {\em IEEE transactions on neural networks and learning systems}, vol.
  28, no. 11, pp. 2660--2673, 2016.

\bibitem{ancona2017towards}
Marco Ancona, Enea Ceolini, Cengiz {\"O}ztireli, and Markus Gross,
\newblock ``Towards better understanding of gradient-based attribution methods
  for deep neural networks,''
\newblock in {\em International Conference on Learning Representations}, 2018.

\bibitem{turbe2022interprettime}
Hugues Turb{\'e}, Mina Bjelogrlic, Christian Lovis, and Gianmarco Mengaldo,
\newblock ``Interprettime: a new approach for the systematic evaluation of
  neural-network interpretability in time series classification,''
\newblock {\em arXiv preprint arXiv:2202.05656}, 2022.

\bibitem{ancona2017unified}
Marco Ancona, Enea Ceolini, Cengiz {\"O}ztireli, and Markus Gross,
\newblock ``A unified view of gradient-based attribution methods for deep
  neural networks,''
\newblock {\em NIPS 2017-Workshop on Interpreting, Explaining, and Visualizaing
  Deep Learning}, 2017.

\bibitem{tan2021icentia11k}
Shawn Tan, Guillaume Androz, Ahmad Chamseddine, Pierre Fecteau, Aaron
  Courville, Yoshua Bengio, and Joseph~Paul Cohen,
\newblock ``Icentia11k: An unsupervised representation learning dataset for
  arrhythmia subtype discovery,''
\newblock in {\em 2021 Computing in Cardiology (CinC)}, 2021.

\bibitem{tan2022physionet}
Shawn Tan, Guillaume Androz, Ahmad Chamseddine, Pierre Fecteau, Aaron
  Courville, Yoshua Bengio, and Joseph~Paul Cohen,
\newblock ``Icentia11k single lead continuous raw electrocardiogram dataset
  (version 1.0),'' 2022,
\newblock \textit{PhysioNet}. https://doi.org/10.13026/kk0v-r952.

\bibitem{jones2020improving}
Yola Jones, Fani Deligianni, and Jeff Dalton,
\newblock ``Improving ecg classification interpretability using saliency
  maps,''
\newblock in {\em 2020 IEEE 20th International Conference on Bioinformatics and
  Bioengineering (BIBE)}. IEEE, 2020, pp. 675--682.

\bibitem{he2016deep}
Kaiming He, Xiangyu Zhang, Shaoqing Ren, and Jian Sun,
\newblock ``Deep residual learning for image recognition,''
\newblock in {\em Proceedings of the IEEE conference on computer vision and
  pattern recognition}, 2016, pp. 770--778.

\end{thebibliography}

\end{document}